
\def \squarebox{$\sqcap\kern-1.5ex\sqcup$} 


\def\IR{{\bf R}}  
\def\ah#1{{\IR^{#1}}} 


\def\word#1{\quad\hbox{#1}\quad}   
\def\Proof#1 {\medbreak\noindent{\bf Proof #1}\quad}
\def\Lemma#1 {\medbreak\noindent{\bf Lemma #1}\quad}
\def\Theorem#1 {\medbreak\noindent{\bf Theorem #1}\quad}

\def\boundary{{\partial\Omega}}
\def\killpsi{r\partial\psi/\partial r}

\magnification\magstep1

\font\bigfont=cmr10 at 20pt

\rightline {NCL-90 TP-16}

\vskip 1in

{\centerline {\bigfont Sources of symmetric potentials}} 

\vskip .5in

\centerline {John W. Barrett}

\vskip.5in
\noindent
\hfill{December 1990}

 \vskip.5in

\beginsection Introduction and main result

Laboratory experiments on gravitation are usually performed with objects
of constant density, so that the analysis of the forces concerns only
the geometry of their shape. In an ideal experiment, the shapes of the
constituent parts will be optimised to meet certain mathematical
criteria, which ensure that the experiment has maximal sensitivity. 

Using this idea, the author suggested an experiment to determine the
departure of the gravitational force from Newton's force law [1]. The
geometrical problem which has to be solved is to find two shapes which
differ significantly, but have the same Newtonian potential.
Essentially, the experiment determines whether the two objects are
distinguishable by their gravitational force. Here, we consider the case
when one of them is a round ball. The result, Theorem 1, establishes a
fact which appeared in numerical simulations, that the second object has
to have a hole in it.

Consider the Laplace equation in $\ah n$, $n\ge 2$, and let the surface
area of a unit sphere in $\ah n$ be $\omega_n$. The fundamental solution
$$
K(x,\xi)=\left\{ \eqalign{
{|x-\xi|^{2-n}\over(2-n)\omega_n} & \word{for} n>2,\cr
{\log|x-\xi|\over2\pi} & \word{for} n=2 \cr
}\right.\eqno(1)
$$
can be regarded as the
potential due to a point source. Let $\Omega$ be a connected open
bounded domain of $\ah n$, with a smooth boundary. Define the Newtonian
potential of $\Omega$ to be
$$
\phi(\xi)=\int_\Omega K(x,\xi)\,dx   \eqno(2)
$$
As is well known this defines a $C^1$ function of $\ah n$ which is
$C^\infty$ on $\ah n\setminus\boundary$ and satisfies
$$
\triangle\phi=\left\{ \eqalign{
1 &\word{on} \Omega\cr
0 &\word{on} \ah n\setminus\overline\Omega\cr}
\right.   \eqno(3)
$$
If $\Omega$ is a solid round ball, or a set of concentric shells, then
on the {\sl outside} of the set $\Omega$, i.e., on the component
$\Sigma$ of $\ah n\setminus\Omega$ which is connected to infinity,
$\phi(\xi)$ is proportional to $K(x,\xi)$, with $x$ the centre of
symmetry. 

Which other sets $\Omega$ share this property? In [1], a
picture of an approximate solution appears, which does not possess
spherical symmetry. I coined the term `monopole' for such sets, on
account of the vanishing of all other multipole moments. The term
`centrobaric' has also been used to denote a more general case of such
solutions to Poisson's equation, where a variable positive density
appears in (3)[2]. The main result presented here is that the hole
which is to be seen in the interior of the example of [1] is a
necessary feature of all asymmetric solutions, in the following manner.

\Theorem 1 Let $\Omega$ be a connected open bounded domain of $\ah n$,
$n\ge 2$, with a smooth boundary, such that $\Sigma=\ah
n\setminus\Omega$ is connected. Let $\phi$ be its Newtonian potential, and
suppose that $\phi$ has the exact form of a point mass on 
$\Sigma$, i.e., 
$$
\phi(\xi)=M\,K(0,\xi)  \eqno(4)
$$
for a constant $M$. Then  $\Omega$ is a solid round ball,
$|x|< const.$\bigbreak  

 The proof
proceeds in two parts, first establishing a bound for the interior
potential, then by applying a rotational version of a method due
to Serrin [3]. Serrin's problem was to consider $\triangle\phi=1$ on a
bounded domain with $\phi=0$ and $\partial\phi/\partial n=const.$ on the
boundary, and ask for the possible shapes for the domain. Serrin
investigated the solutions to this free boundary problem by exploiting the
invariance of Poisson's equation, the source term and the boundary
condition by reflection in a plane. The plane was adjusted so that a
reflected portion of the boundary touched the boundary itself. In our
case, there are not enough reflection planes which respect the form (1)
of the potential on the outside of $\Omega$. The appropriate symmetries
to use are rotations and homotheties (linear scalings) about the origin.
Thus one could shrink and rotate the figure so that its boundary touched
the original boundary at a point. Lemma 3 is a slightly modified version
of this idea, where the second figure is replaced by a ball.

\beginsection Details

Let us assume that $\Omega, \phi,\Sigma$ are defined as in Theorem 1. 
We are interested in comparing the interior potential with the point
mass potential (4).

\Lemma 2 
Let $\phi$ obey equation (4) everywhere on $\Sigma$, which, for the
purposes of this lemma need not be connected. Then it follows that 
$$
\phi(\xi)> M\,K(0,\xi)   \eqno(5)
$$
for $\xi\in\Omega\setminus\{0\}$, and that $\Omega$ is
star-shaped about 0.\bigbreak

\Proof : Note first that (4) implies that $0\in\Omega$. Consider the
function
$$
\psi(\xi)=\phi(\xi)-M\,K(0,\xi)    \eqno(6)
$$
on $\ah n\setminus0$. Then $\psi=0$ and $\nabla\psi=0$ on $\boundary$.
Let $r=|\xi|$ denote the radial coordinate in spherical coordinates. The
function $\killpsi$ defined on the set 
$\overline\Omega\setminus\{0\}$ has some interesting properties;
considering its behaviour is a mathematical counterpart to decomposing
the source $\Omega$ into concentric shells. One has that $\killpsi\in
C^0(\overline\Omega\setminus\{0\}) \cap C^2(\Omega\setminus\{0\})$,
$\triangle\bigl(\killpsi\bigr)=2$ on $\Omega\setminus\{0\}$, $\killpsi=0$
on $\boundary$, and 
$$
r{\partial\psi\over\partial r}=r\left({\partial\phi\over\partial
r}-{M\over\omega_n}r^{1-n}\right)<0                            \eqno(7)
$$
for $r$ sufficiently close to 0, $0<r<\epsilon$, say. Thus by applying
the maximum principle to $\Omega$ with the ball of radius $\epsilon$
removed, the inequality (7) extends to the whole of
$\Omega\setminus\{0\}$. 

Now consider the values of $\psi$ along a straight line segment joining
$p\in\Omega$ to 0. The inequality (7) implies that the whole of the
segment must lie in $\Omega$ and that $\psi>0$ in $\Omega$. 
\squarebox\medbreak

\Lemma 3 Let $\phi$ obey (4) on $\Sigma=\ah n\setminus\Omega$, for an
$\Omega$ as described in Theorem 1. If, in addition, (5) is
obeyed on $\Omega\setminus\{0\}$, then the boundary, $\boundary$,
is a sphere, $|x|=const.$\medbreak

\Proof : Suppose $\boundary$ is not a sphere, so that the
maximum and minimum distance of its points from the origin satisfy
$r_{max}>r_{min}>0$. Let $m\in\boundary$ be a point at which a minimum
distance is attained. The open ball $B$ of radius $r_{min}$ centred at 0
has boundary $\partial B$ which touches $\boundary$ at $m$. Denote its
Newtonian potential by $\phi_B$,  and set
$$
u(\xi)={M\over vol(B)}\phi_B(\xi)-\phi(\xi).  \eqno(8)
$$
The scaling has been chosen so that $u=0$ on $\Sigma$. Now,
$B\subset\Omega$ and $vol(B)<M=vol(\Omega)$. Thus $\triangle
u=M/vol(B)-1>0$ on $B$, $u=0$ on $\boundary\cap\partial B$, and $u<0$ on
$\Omega\cap\partial B$, by virtue of (5). Using the maximum principle,
$u<0$ on $B$. As is common to Serrin's method, there is a point $m$
on the boundary at which $u=\partial u/\partial r=0$.
 The Hopf lemma (see [4]) applied to a small ball $C\subset B$,
$m\in\overline C$, yields a contradiction, unless $u\equiv0$, in which
case $\boundary$ and $\partial B$ coincide. This contradicts the
supposition $r_{max}>r_{min}$, and so the boundary is a
sphere.\quad\squarebox

\Proof {of Theorem 1}  This follows from lemmas 2 and
3.\squarebox\medbreak

Asymmetric monopoles with holes (i.e., $\Sigma$ not connected) escape
Theorem 1 because (4) is not obeyed in the interior holes. In
fact, the conclusion of Lemma 2 means that it is fruitless to
enforce (4) on an interior hole, because then $\Omega$ cannot
have any holes.

\beginsection References

\frenchspacing
\item{[1]} J.W. Barrett: The Asymmetric Monopole and non-Newtonian
Forces.\quad  Nature {\bf341} 131-132 (1989)

\item{[2]}  W.D. MacMillan: Theory of the Potential. New York, 1930

\item{[3]} J. Serrin: A Symmetry Problem in Potential Theory.\quad  Arch.
Ration. Mech. {\bf43} 304-318 (1971)

\item{[4]} B. Gidas, Wei-Ming Ni and L. Nirenberg: Symmetry and Related
Properties via the Maximum Principle.\quad Comm. Math. Phys. {\bf68},
209-243 (1979)

\bye